\documentclass[superscriptaddress,twocolumn,prb]{revtex4}
\usepackage{graphicx}
\usepackage{dcolumn}
\usepackage{bm}
\usepackage{color}
\usepackage{ulem}
\usepackage{amsmath}
\usepackage{amsfonts}
\usepackage{amssymb}
\usepackage{tikz}
\usepackage[colorlinks=True,linkcolor=blue,anchorcolor=red,citecolor=blue,CJKbookmarks=true]{hyperref}

\usepackage[final]{changes}

\begin{document}

\title{Phase transitions and topological properties of the 5/2 quantum Hall states with strong Landau-level mixing}
\author{Wenchen Luo}
\affiliation{School of Physics and Electronics, Central South University, Changsha, China 410083}
\author{Wei Zhang}
\affiliation{School of Physics and Electronics, Central South University, Changsha, China 410083}
\author{Yutao Hu}
\affiliation{School of Physics and Electronics, Central South University, Changsha, China 410083}
\author{Hao Wang}\email{wangh@sustech.edu.cn}
\affiliation{Shenzhen Institute for Quantum Science and Engineering,
Southern University of Science and Technology, Shenzhen, China 518055}
\affiliation{International Quantum Academy, Shenzhen, China 518048}
\affiliation{Guangdong Provincial Key Laboratory of Quantum Science and Engineering, Southern University of Science and Technology, Shenzhen, China 518055}

\date{\today }

\begin{abstract}
We numerically study a 5/2 fractional quantum Hall system with even number of electrons using the exact diagonalization where both the strong Landau level (LL) mixing and a finite width of the quantum well have been considered and adapted into a screened Coulomb interaction. With the principal component analysis, we are able to recognize a compressible-incompressible phase transition in the parameter space made of the magnetic field and the quantum well width by the competition between the first two leading components of the ground states wave functions, which is consistent with the low-lying spectral feature and previous works in the odd-electron system. In addition, the presence of the subdominant third component suggests an incompressible transition occurring as the LL-mixing strength grows into a certain parameter region associated with the ZnO experiments. We further investigate the strongly LL-mixed phase in this emerging region with the Hall viscosity, wave function overlaps, and the entanglement spectra. Results show it can be well described as a particle-hole symmetrized Pfaffian state with the dual topological properties of the Pfaffian and the anti-Pfaffian states.
\end{abstract}

\maketitle

\section{Introduction}

The nature of the even denominator fractional quantum Hall effect (FQHE) is still unclear in many aspects\cite{odfqhe}. In wide or double quantum wells, even denominator FQHE can be explained by the Halperin state \cite{halperin}. But for those FQHEs with nonabelian excitations, an understanding for their nature is still in puzzle. For 5/2 FQHE, the Pfaffian \cite{mr} or anti-Pfaffian wave function \cite{apf} is proposed to be the most promised candidate. However, either the numerical calculation of wave function overlaps or the thermal conductance experiment \cite{thermal} challenges these trial wave functions. Some theoretical approaches try to explain the thermal conductance experiment by modeling the ground state with Pfaffian-anti-Pfaffian domains \cite{pfapfdomain} or by considering the partial equilibration of the thermal transport \cite{nonequilibrum}. Other trial wave functions are also proposed in the spirit of the composite fermion paring \cite{pairing,son}. However, they are either energetic unfavorable or unmatched with numerical overlap calculations. Nevertheless, the topologically nontrivial feature of the $5/2$ FQHE, which is experimentally confirmed by the Majorana mode with $1/2$ thermal Hall conductance, should be recognized.

In studying the FQHE of a realistic two-dimensional electron gas (2DEG), Landau level (LL) mixing is an important effect to be considered. To scale the strength of the LL-mixing, ones have defined a dimensionless parameter $\kappa=E_C/ E_{LL}$ where $E_C= e^2/\epsilon \ell$ is the Coulomb interaction strength and $E_{LL} =\hbar \omega_c$ is the LL gap with the cyclotron frequency $\omega_c=|e|B/m^*$ as $e$ for the electron charge, $B$ for the magnetic field, and $m^*$ for the effective mass. If $\kappa \rightarrow 0$, the LLs are considered infinitely gapped and thus the LL-mixing is negligible.

For the GaAs quantum well, considering the material parameters and the magnetic field used in experiments, the LL-mixing there has a moderate strength with $\kappa \sim 1$, which is strong enough to affect the nature of the FQHE state. It has been argued that the perturbative theory would still be feasible and the induced three-body interaction will break the particle-hole (PH) symmetry \cite{perturbation}. Given that the (anti)Pfaffian state is the unique zero-energy state of some three-body interaction in a half-filled system, the 5/2 FQHE state is thus related to a Pfaffian or an anti-Pfaffian state when the LL-mixing is taken into account. However, in the ZnO/MgZnO heterostructure \cite{falson,falson2, joe_more} or black phosphorene \cite{bpfqheexp} the effective mass of the electron is about one magnitude higher than that in GaAs quantum well, leading to an extremely large $\kappa \gg 1$. There, the LL-mixing is so strong that the perturbative theory could not be available. More importantly, the strong LL-mixing has been argued possible to recover the particle-hole symmetry \cite{feldman}, which will be reclaimed in this work. It is reasonable
to expect that the ground states of the $5/2$ FQHEs in ZnO quantum well should be biased from that in GaAs.

For a strongly LL-mixing system with a finite filling (for instance $\nu=5/2$), the electrons are distributed in far more than three LLs, causing a huge Hilbert space. The computer resource for such a large system would be exhausted in
most conventional approaches. A substitutive way to study the strongly LL-mixed system is to project the 2DEG into a single relevant LL with the effectively screened Coulomb interaction in the random phase approximation (RPA) \cite{luo1,luo3,rpa}, of which the Hilbert space is sufficiently small for the exact diagonalization scheme. The screened Coulomb potential containing the information of all the other LLs can be used to interpret the stability of the $5/2$ FQHE \cite{luo1,luo4}, which agrees with the experimental results \cite{falson,falson2}. In our present work on the strongly LL-mixed 5/2 fractional quantum Hall system with a finite width of the quantum well, we adopt the same screening treatment for the electron-electron interaction to construct an effective model Hamiltonian. With the numerical spectra and principal component analysis (PCA) \cite{pca} on the ground state wave functions, we obtain a phase diagram in the parameter space consist of the magnetic field and the width of the quantum well with several phase transitions. An incompressible phase with a large LL-mixing strength can be associated to parameters of the ZnO samples and physical properties of its ground state are investigated through the calculations of the Hall viscosity, wave function overlaps, and the entanglement spectra \cite{entangle}.

\section{Model Hamiltonian}

We consider a 5/2-filling FQHE system with the 2DEG extended a bit in the perpendicular ($z$) direction due to the finite width of the quantum well. For simplicity, the confinement potential of the quantum well is assumed to be an infinity square well. Thus, the $z-$component of the wave function is given by $\phi_m(z)=\sqrt{2/ L^{}_z}\sin(m \pi z/L_z)$ with the quantum number $m$ indexing the band and $L_z$ for the width of the quantum well. The full wave function of the 2DEG takes a two-part form as $\Psi\left(\mathbf{r},z\right) =\phi_m(z) \psi\left(\mathbf{r}\right)$, where $\mathbf{r}$ is the in-plane coordinate. On the torus, the in-plane part of the wave function has the form of $\psi_{n,i}\left(\mathbf{r}\right)$ in the Landau gauge with $n$ for the LL\ index and $i$ for the guiding center index. On a sphere, the in-plane part $\psi$ should be replaced by the wave function on the zero-width sphere, and 
$i$ for the angular momentum.

In the toroidal geometry, The many-body Hamiltonian of a Coulomb interaction system is given by
\begin{eqnarray}
&&H=\sum_{n,i,\sigma}E^{}_{n,\sigma }c_{n,i,\sigma }^{\dag }c^{}_{n,i,\sigma}+
\frac12\sum_{\sigma,\sigma^{\prime}}\sum_{\substack{n^{}_1,\ldots,n^{}_4  \\
i^{}_1,\ldots,i^{}_4}} V_{\left(s\right),i^{}_1,i^{}_2,i^{}_3,i^{}_4}^{n^{}_1,n^{}_2,
n^{}_3,n^{}_4}  \notag \\
&&\times\, c_{n^{}_1,i^{}_1,\sigma}^{\dag}c_{n^{}_2,i^{}_2,\sigma^{\prime
}}^{\dag }c^{}_{n^{}_3,i^{}_3,\sigma^{\prime}}c^{}_{n^{}_4,i^{}_4,\sigma},
\end{eqnarray}
where the first term is kinetic energy with the summation over all LL indices $n$, guiding center indices $i$, and spin indices $\sigma$. Symbols $c$ ($c^\dag$) stands for electron annihilation (creation) operator. As to the electron-electron interaction term, considering the strong LL-mixing, we treat it with the screening scheme as modeled in Ref. [\!\!\citenum{luo4}]. The electron distribution is the same as the noninteracting picture so that we can concentrate ourselves on the highest occupied LL, in which the electron-electron interaction is screened by all the virtual processes between the LLs with different filling in the RPA. We believe that this screening scenario is a good approach to describe the strongly LL-mixed system although some of the correlations are neglected. Under such treatment, we can constrain the system in the half-filled $n=1$ LL. If all electrons can be regarded fully spin polarized due to the large magnetic field \cite{luo1}, the Hamiltonian is reduced to
\begin{equation}
H=\frac12 \sum_{i^{}_1,\ldots,i^{}_4}
V^{n,s}_{i^{}_1,i^{}_2,i^{}_3,i^{}_4} c_{i_1}^{\dag}c_{i_2}^{\dag }c_{i_3} c_{i_4},
\label{hamiltonian}
\end{equation}
where the screened Coulomb interaction matrix element has the form of
\begin{eqnarray}
V^{n,s}_{ i^{}_1,i^{}_2,i^{}_3,i^{}_4} &=&
\frac{e^2}{\epsilon \ell} \frac{2}{\pi N^{}_s}\overline{\sum_{\mathbf{q}}}
\delta _{i_{1},i_{4}+q_{y}\ell ^{2}}^{\prime }\delta
_{i_{2},i_{3}-q_{y}\ell ^{2}}^{\prime }e^{iq_{x} \ell^2 \left( i_{3}-i_{1}\right) } \notag
\\
& \times& e^{-\frac{q^2 \ell ^2}{2}} \left[ L_{n }\left( \frac{q^{2}\ell ^{2}}{2}\right) \right]^2
V_s(q).
\end{eqnarray}
In the above expression, $\mathbf{q}$ is the discrete in-plane momentum, $N_s$ is the LL degeneracy, $\delta^{\prime }$ includes the periodic boundary condition, $L_n(x)$ is a Laguerre polynomial, and $\overline{\sum}$ means that the $q=0$ term is excluded in the summation.

As we consider only the first band at $m=1$, the effective Coulomb potential $V_s(q)$ is given by
\begin{equation}
V_s(q) = \int_0^{\infty }\frac{ 32\pi^4 dq^{}_z\ell}{\epsilon^{}_s\left(
q,q^{}_z\right) \left( q^2+q_z^2\right) \ell^2} \frac{
1-\cos \left( q^{}_zL^{}_z\right) }{\left( 4\pi^2q^{}_zL^{}_z-q_z^3L_z^3\right)^2}.
\end{equation}
Such dressed Coulomb potential is obtained by the pure Coulomb potential divided by the static dielectric function $\epsilon_s(q,q_z)$. In calculating the dielectric function, we use the noninteracting retarded density-density response function $\chi _{nn}^{0}\left( q ,q_{z},\omega  \rightarrow 0 \right) $ in the RPA 
in the static limit for simplicity. The dielectric function is then given by
\begin{equation}
\epsilon^{}_s\left(q\mathbf{,}q^{}_z\right) =1-\frac{4\pi e^2}{\left(
q^2+q_z^2\right)\epsilon}\chi_{nn}^0\left(q\mathbf{,} q^{}_z\right)
\end{equation}
with
\begin{eqnarray}
\chi _{nn}^{0}\left( q ,q_{z}\right) & =& \chi _{nn}^{0}\left( q%
\mathbf{,}q_{z}\mathbf{,}\omega \rightarrow 0\right)  \notag \\
&=&\frac{1}{2\pi \ell ^{2}L_{z}}\sum_{\sigma
}\sum_{m_{1},n_{1}}\sum_{m_{2},n_{2}}\left\vert
G_{m_{2},n_{2}}^{m_{1},n_{1}}\left( \mathbf{q},q_{z}\right) \right\vert ^{2}
\notag \\ &\times&
\frac{\nu _{\left( m_{1},n_{1}\right) ,\sigma }-\nu _{\left(
m_{2},n_{2}\right) ,\sigma }}{E_{\left( m_{1},n_{1}\right) ,\sigma
}-E_{\left( m_{2},n_{2}\right) ,\sigma }},  \label{chi}
\end{eqnarray}%
where $m_i$ and $n_i$ mark the band and the LL indices, $E_{\left( m_{1},n_{1}\right) ,\sigma }$ and $\nu _{\left( m_{1},n_{1}\right) ,\sigma }$ are the kinetic energy and the filling factor of the LL $\left( m_{1},n_{1}\right) $ with spin $\sigma $, respectively. The form factor is written as
$G_{m_{2},n_{2}}^{m_{1},n_{1}}\left( \mathbf{q},q_{z}\right)
=iq_{z}L_{z} F_{n_{1},n_{2}}\left( -\mathbf{q}\right)
g_{m_{1},m_{2}}\left( -q_{z}\right)$
with the functions%
\begin{eqnarray}
&&F_{n,n^{\prime }}\left( \mathbf{q}\right) =\frac{\sqrt{\min \left(
n,n^{\prime }\right) !}}{\sqrt{\max \left( n,n^{\prime }\right) !}}e^{-\frac{%
q^{2}\ell ^{2}}{4}}L_{\min \left( n,n^{\prime }\right) }^{\left\vert
n-n^{\prime }\right\vert }\left( \frac{q^{2}\ell ^{2}}{2}\right)
\notag \\ && \times
\left[ \frac{\mathtt{sgn}\left( n-n^{\prime }\right) q_{y}\ell
+iq_{x}\ell }{\sqrt{2}}\right] ^{\left\vert n-n^{\prime }\right\vert }
\label{ffunction} \\
&&g_{m_{1},m_{2}}\left( q_{z}\right) =\left[ 1-e^{iq_{z}L_{z}}\cos \left(
m_{1}-m_{2}\right) \pi \right]
 \\ && \times
\left[ \frac{1}{\left( m_{1}-m_{2}\right) ^{2} \pi^2 -
q_{z}^{2}L_{z}^{2} }-\frac{1}{\left( m_{1}+m_{2}\right) ^{2}\pi^2-
q_{z}^{2}L_{z}^{2}}\right] .  \notag
\end{eqnarray}
The expression of Eq. (\ref{chi}) indicates a summation over all LLs. In practise, it is necessary to set a truncation since the high-energy LLs contribute little to the studied system.

In the spherical geometry, the Hamiltonian can also be written in a form similar to Eq. (\ref{hamiltonian}) by using the Haldane's pseudopotential
\begin{equation}
V_{i}^{ s } =
\int_0^\infty \frac{dq^2\ell^2}{\pi}  \left[ L_{n}\left( \frac{q^{2}\ell ^{2}}{2}\right) %
\right] ^{2}e^{-q^{2}\ell ^{2}}L_{i}\left( q^{2}\ell ^{2}\right)
V_s(q),
\end{equation}
where we approximately use the planar pseudopotential and both the thickness and the screening corrections have been included in $V_s(q)$.

As shown in the above effective Hamiltonians, both the strength of the magnetic field and the width of the quantum well will soften the Coulomb interaction and thus significantly modify the properties of the strongly correlated system. Topological phase transitions might occur \cite{luo3} given continuous variations in the parameter space made of $B$ and $W$, which is the half-width of the wave function in the $z$ direction and $W=L_z/1.5$ for an infinity square well. Such possible phase transitions and the topological features of the corresponding quantum states can be explored through the calculated energy spectra and state wave functions as we numerically solve the model Hamiltonians using the exact diagonalization. Since different topological states on the sphere may correspond to different spherical shifts, thus different Hilbert spaces, it could be difficult to probe a possible topological phase transition with continuously varying parameters \deleted{for a fixed system} in the spherical geometry. \replaced{Therefore}{Instead}, in our analysis we mainly take the toroidal geometry \cite{haldane, book}. The parallelogram unit cell of the torus is framed by two basis vectors $\vec{L}_1$ and $\vec{L}_2$ with the aspect ratio $|\tau|=L_1/L_2$ and the aspect angle $\theta$ between them. Within the cell, there are  $N_e$ electrons and $N_s$ magnetic flux quanta with $N_s=2N_e$ corresponding to the half-filled second LL. Several state wave functions in the spherical geometry at specified spherical shifts are also studied for comparison purpose.

\section{Phase transitions}

In a previous work \cite{luo4}, strongly LL-mixed systems with odd number of electrons on the torus have been studied in the $(B,W)$ parameter space. The compressible-incompressible phase transition is detected there by monitoring the variation of the excitation energy gap. The 2DEG is found generally incompressible in the $B\cdot W > 30$ T$\cdot$nm region and the phase diagram can be used to explain the missing or appearance of the experimental observation of the 5/2 FQHE in different ZnO samples \cite{falson2,joe_more}. However, considering the underlying pairing nature of the even denominator FQHE, a study on an even-electron system would be more relevant and may provide more fundamental information. Different from an odd-electron system where the ground state is non-degenerate and a well-defined excitation gap can serve as the incompressibility probe, an even-electron system on the torus can have its ground state locating ``degenerately'' at three characteristic pseudomomentum sectors $(N_e/2,0), (0,N_e/2)$ and $(N_e/2,N_e/2)$ (with the two-fold center of mass degeneracy subtracted) due to the topology nature of the Pfaffian state \cite{peterson} and the complete degeneracy is lifted in a finite-sized system. Thus, for the even-electron system in our consideration it is in doubt whether we still can determine an incompressible gap to detect the phase transition. Nevertheless, we can first observe the low-lying energy spectra of the system at distinct phase regions to gain a general impression.

\subsection{Energy spectra}

\begin{figure}[htp]
  \centering
  \includegraphics[width=8.4cm]{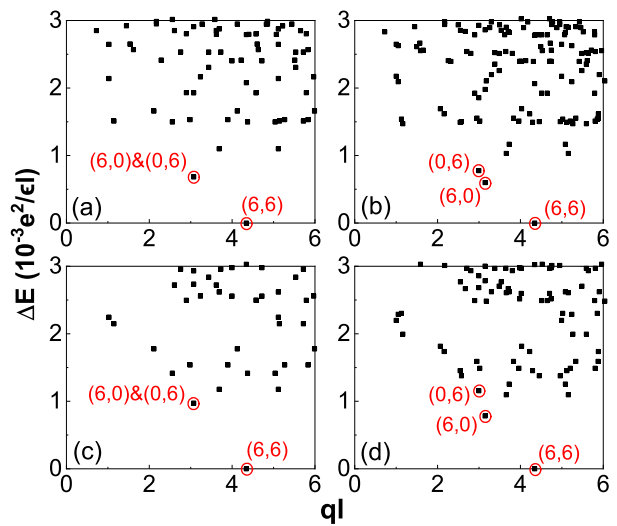}
  \caption{(Color online) The low-lying excitation spectra of an $N_e=12$ system on the torus with a square unit cell for (a) and (c) but a rectangular cell of the aspect ratio $|\tau|=0.95$ for (b) and (d). The parameter set ($B=9.6$ T,$W=5$ nm) used in (a) and (b) stands for the experimental ZnO sample $a$ while ($B=7.2$ T,$W=5$ nm) in (c) and (d) for sample $b$. The ground states at three characteristic pseudomomentum sectors have been labeled and marked in the spectra.}
  \label{fig1}
\end{figure}

To get a close comparison with the experimental results of the ZnO quantum well \cite{falson2}, we investigate the excitation spectra of the even-electron systems at two selected parameter sets. One is corresponding to the incompressible 2DEG in the sample $a$ ($B=9.6$ T and $W=5$ nm) and the other ($B=7.2$ T and $W=5$ nm) for the compressible 2DEG in sample $b$. The material parameters are adopted from the ZnO for the effective mass $m^*=0.3m_e$ with electron mass $m_e$ and the dielectric constant $\epsilon=8.5$.

In Figure \ref{fig1}, we show the low-lying spectra of an $N_e=12$ system with a square unit cell and a more general rectangular cell at $|\tau| =0.95$. As seen in Figs. 1(a) and 1(b) for the incompressible sample $a$, the lowest three states remain staying at the three characteristic sectors of (0,6),(6,0) and (6,6), alike the Pfaffian state on the torus. If we treat them collectively as a three-fold manifold gaped from other higher energy states, we note this separation gap and the splitting band width among them are sensitive to the cell geometry. However, for the compressible sample $b$, this general feature of characteristic three-fold manifold is missing though the (6,6) state remains as the global ground state. In  Fig. 1(c), the lowest (0,6) state and the lowest (6,0) state (degenerate pair due to the additional symmetry of the square cell) are much closer to the upper states in comparing with the incompressible sample case. In Fig. 1(d) with a rectangular cell, the lowest (0,6) state is lifted from the lowest (6,0) state and no longer holds in the characteristic three-fold manifold since some states at other sectors have lower energies.

These spectral observations suggest the incompressible phase can be qualitatively associated with the form of the characteristic three-fold manifold. However, due to the finite size effect it is ambiguous to define an energy gap between this manifold and a higher excitation level. Moreover, this manifold is only valid for the ground state with pfaffian-like properties. These issues limit our ability of using the energy spectrum alone to quantitatively detect the compressible-incompressible transition for an even-electron system. Instead, we turn to use calculated wave functions as the probe for the phase transition. Considering the lack of explicit knowledge for the quantum states at different phase regions, we exploit the PCA method to analyze the system  as described in the next subsection.

\subsection{Principal component analysis}

As an unsupervised machine learning approach, PCA has been applied to study quantum phases in several FQHE systems \cite{pca}. There, distinct principal components (PCs) are extracted directly from the parameterized wave functions using singular value decomposition technique. By tracing the evolution and switching of the predominant PCs, ones are able to categorize different quantum phases and detect their transitions.

\begin{figure}[htp]
  \centering
  \includegraphics[width=8.5cm]{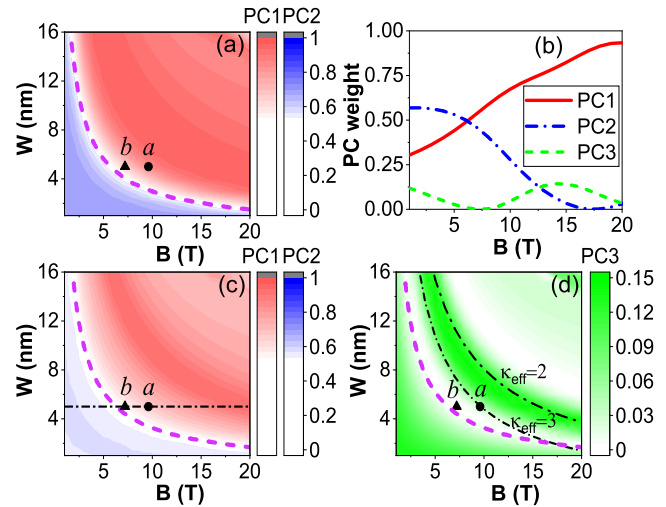}
  \caption{(Color online) Weight distributions of the leading PCs in the ($B$,$W$) parameter space for the manifold states of an $N_e=12$ system on the torus with a square unit cell. The plot (a) stands for two leading PCs of the ground state at (0,6) pseudomomentum sector where the dashed (purple) curve sketches the compressible-incompressible phase boundary by the switching of the two PCs and the locations of the experimental ZnO samples $a$ and $b$ are marked. The plot (b) shows the weights of three leading PCs as a function of $B$ for the ground state at (6,6) sector when the width is fixed to  $W=5$ nm. The plot (c) stands for the first two PCs of the (6,6) ground state with the dot dash line corresponding to the plot (b). The plot (d) stands for the third leading PC of (6,6) ground state with its largest distribution inside the incompressible region mostly falling in the range of $\kappa_{\text{eff}} \in [2,3]$.}\label{fig2}
\end{figure}

In our PCA practise, considering the spectral features and the importance of the three lowest states in the characteristic sectors $(N_e/2,0), (0,N_e/2)$ and $(N_e/2,N_e/2)$, the input data are taken from the ground states wave functions at these characteristic sectors and the investigated parameter space spans in a range of $B\in[1,20$ T] and $W\in[1,16$ nm] in adapting to the ZnO experiments. For simplicity, we consider the LL-mixed system with a square unit cell and an even number of electrons up to $N_e=12$. Since the $(N_e/2,0)$ state and the $(0,N_e/2)$ state are degenerate due to the square symmetry, we discuss only the $(0,N_e/2)$ and $(N_e/2,N_e/2)$ states in the following. By monitoring the weights of PCs in the considered parameter space, we note only \added{a} few PCs [two for $(0,N_e/2)$ state and three for $(N_e/2,N_e/2)$ state] exhibit significant non-zero presence with their sum weight near unity and other PCs are negligible. Therefore, we focus these leading PCs in our analysis. By tracing the relative evolution of the leading PCs, we are able to draw a phase diagram of an example $N_e=12$ system as shown in Fig. \ref{fig2}.

For the lowest $(0,6)$ state, there are two leading PCs competing and their weights switch along a curved boundary (dashed purple line) in the parameter space as shown in Fig. 2(a), indicating the occurrence of a phase transition. We note that an earlier work \cite{luo4} with an odd-electron ($N_e=11$) system has obtained a similar phase diagram where the compressible and incompressible phase regions are identified through the incompressible gap. Mapping with those results, we expect the parameter region in our diagram with the large $B$ and $W$, i.e., the first PC dominated region, would correspond to an incompressible phase while the phase at the opposite side would be compressible. The further evidences are shown as we mark two experimental samples with their corresponding parameters in the diagram. Although both samples seem locating in the assumed incompressible side, the sample $b$ is already much close to the boundary. One might suspect that the sample $b$ actually falls in the transition zone to be more compressible especially at finite temperature, which is also demonstrated by its low-lying spectra. 

For the $(6,6)$ ground state, there are existing three leading PCs as one example is shown in Fig. 2(b) for the wave functions at the width $W=5$ nm. The competition and switching between the first two PCs provide us with a similar phase diagram for the compressible-incompressible phase transition as shown in Fig. 2(c) while the subdominant third PC presents some extra feature. Within the incompressible zone, we note there is a narrow region near the compressible-incompressible phase boundary where the third PC emerges in companying with the suppression of the second PC. If we define a rescaled LL-mixing parameter $\kappa_{\text{eff}}= e^{-L_z/\ell} E_C /E_{LL}$ in considering the effect of the finite width of the quantum well, this particular region mostly locates within the range of $\kappa_{\text{eff}}\in [2,3]$ as shown in Fig. 2(d). Since the composition of the PCs at this region is different from that with less $\kappa_{\text{eff}}$, a phase transition is suspected as the LL-mixing strength of an incompressible system grows to the $\kappa_{\text{eff}}\in [2,3]$ range. We note some earlier experimental and numerical work \cite{luo3,dopedgaas} for a strong LL-mixed system with zero-width quantum well have suggested a possible phase transition occurring within the incompressible range $\kappa\in [2,3]$. Our discovery with the subdominant PC support their results.

\section{Incompressible quantum state with strong LL-mixing}
For the incompressible phase at small LL-mixing, a few of earlier theoretical and numerical works have proposed it to be a Pfaffian or an anti-Pfaffiant state due to perturbative three-body interactions. At large LL-mixing where the perturbation theory is invalid, our treatment in previous sections with a screened Coulomb interaction suggests an incompressible phase transition emerges as the growing LL-mixing strength approach the $\kappa_{\text{eff}}\in [2,3]$ range. Even if we leave the incompressible region with less $\kappa_{\text{eff}}$ as a transition zone from the small perturbative LL-mixing phase, there absents knowledge for the incompressible states within the emerged phase region $\kappa_{\text{eff}}\in [2,3]$. In this section, we will investigate the physical properties of the quantum states in this region. Note the incompressible sample $a$ of the ZnO experiment \cite{falson2} locates within this region, in the following discussion we choose the corresponding parameter set ($B=9.6$ T and $W=5$ nm) as the representative case for a better understanding of the experiments.

\subsection{Hall viscosity}

Hall viscosity, which is defined as $\eta=\frac{1}{2}\bar{s}\bar{n}\hbar$ with the average particle density $\bar{n}$ and the average orbital spin per particle $\bar{s}$, has been used to characterize the topological property of a FQHE state\cite{hallvis}. It is closely related to the spherical shift $S$, which is another topological probe defined on the sphere\cite{shift}, by relation of $S=2\bar{s}$. On the torus, given the geometry is parameterized by a complex number $\tau=\tau_1+i\tau_2=|\tau|e^{i\theta}$, the mean orbital spin $\bar{s}$ (as well as the relevant $\eta$ and $S$) 
can be calculated through the Berry phase of a considered many-body state $\Psi$ when it adiabatically varies around an enclosed path in the geometric parameter space. For a circular path around a center $\tau_0$ and with a radius $\rho_0 (<\tau_{20})$, it gives:
\begin{equation}
\bar{s}=\frac{1}{2}+\frac{1}{2\pi}\mbox{Im}\int\int d\tau_1 d\tau_2 \left\langle \frac{\partial\Psi}{\partial\tau_1}\right| \left. \frac{\partial\Psi}{\partial\tau_2} \right\rangle/D,
\label{obitspin}
\end{equation}
with the normalization factor $D=N_e[\frac{1}{\sqrt{1-(\rho_0/\tau_{20})^2}}-1]$. In our numerical study, the integral in the second term of the Eq. (\ref{obitspin}) is evaluated by evenly discretizing the circular path into $M$ pieces and the result then can be counted as a sum of the normalized local Berry curvature as $\sum_{k=1}^{M}B_k$. The topological nature of an incompressible state requires the calculated orbital spin $\bar{s}$ (or shift $S$) converges to a certain rational number for a small $\rho_0$ and a large $M$, regardless of the choice of $\tau_0$, e.g., $\bar{s}=3/2$ for Pfaffian state and $\bar{s}=-1/2$ for anti-Pfaffian state.

\begin{figure}[htp]
\centering
\includegraphics[width=8.5cm]{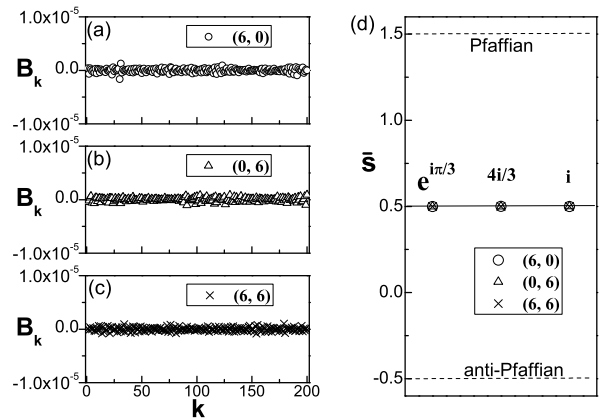}
\caption{The local Berry curvature for the ground states of the $N_e=12$ Coulomb system at pseudomomenta (a) (6,0), (b) (0,6), and (c) (6,6). In the calculation we used 200 discrete steps along a circular path with its center $\tau_0=i$ (square geometry) and radius $\rho_0=0.001$. The resultant $\bar{s}$ around different circular paths are collected in (d) where their corresponding centers $\tau_0$ are labeled and two dashed lines mark the expected values for a Pfaffian state and an anti-Pfaffian state as a comparison.}
\label{hallvs}
\end{figure}

We take Hall viscosity as our first probe to the topological property of the incompressible state with strong LL-mixing. For a system associated with the representative parameter set, we focus on the three lowest states in the characteristic three-fold manifold. The mean orbital spin of the ground state at each characteristic pseudomomentum sector is calculated and the results for an example $N_e=12$ system with $M=200$ and $\rho_0=0.001$ are shown in Fig. \ref{hallvs}. Other smaller systems is found owning quantitatively similar features. 

As presented in Figs. \ref{hallvs}(a)-(c), the local Berry curvatures are homogeneously close to zero in the whole integral region, resulting a vanished sum for the Berry phase. The smooth distribution of the local Berry curvature without dramatic fluctuations indicates that the investigated state carries a stable topological order. Furthermore, the consequent $\bar{s}$ are plotted in Fig. \ref{hallvs}(d) for different integral paths, indicating a universal convergence to $\bar{s}=1/2$, which is neither the expected value for the Pfaffian state nor for the anti-Pfaffian state. Note that a particle-hole symmetric state at half-filling has $\bar{s}=1/2$, our numerical results suggest the investigated incompressible state would more likely be a particle-hole (PH) symmetric state rather than the PH-asymmetric Pfaffian or anti-Pfaffian state. One would not be surprised for the PH symmetry of the ground state as confirmed by the above calculation since the screened Coulomb potential is essentially a two-body interaction. At strong LL-mixing, we believe this RPA renormalized two-body interaction is sufficient enough to make the system regain the PH symmetry which could be broken in the appearance of perturbative three-body interactions in less LL-mixing cases.

\subsection{Wave function overlap}

Considering the alike spectral feature with the same characteristic three-fold manifold as the Pfaffian ($|pf \rangle$) or anti-Pfaffian ($|apf \rangle$) state and the PH-symmetric nature revealed through the Hall viscosity calculation, it is reasonable to construct a PH-symmetrized Pfaffian state ($|spf \rangle$) as a possible candidate state to describe the incompressible system at large LL-mixing. Such symmetrized state is an equal weight superposition of a Pfaffian state and its particle-hole flipped state ($\overline{|pf \rangle}$) in the form of $|spf \rangle= (|pf \rangle \pm \overline{|pf \rangle})/\sqrt{2}$. For a 5/2 FQHE systems with neglected LL-mixing effect, numerical studies showed $|spf \rangle$ can have much larger \deleted{(nearly unitary)} overlap with the system ground state comparing with the (anti-)Pfaffian state. For the strong LL-mixing system, we carry out the same exam with the wave function overlap. The (anti-)Pfaffian states used in the calculation are generated as the densest zero-energy ground state of a three-body pseudopotential \cite{3bpp}.

\begin{table}
\renewcommand\arraystretch{1.5}
\scalebox{1.}{
\begin{tabular}{c|c|c|c}
 \multicolumn{4}{c}{Overlaps}\\
\hline  & $| \langle pf |GS \rangle |$ & $|\langle apf |GS \rangle |$ & $|\langle spf |GS \rangle |$ \\
\hline $\left(\frac{N_e}{2},0\right), N_e=10$&0.797&0.797&0.973\\
\hline $\left(0,\frac{N_e}{2}\right), N_e=10$&0.797&0.797&0.973\\
\hline $\left(\frac{N_e}{2},\frac{N_e}{2}\right), N_e=10$&0.73&0.73&0.989 \\
\hline $\left(\frac{N_e}{2},0\right), N_e=12$&0.76&0.76&0.958\\
\hline $\left(0,\frac{N_e}{2}\right), N_e=12$&0.76&0.76&0.958\\
\hline $\left(\frac{N_e}{2},\frac{N_e}{2}\right), N_e=12$&0.776&0.776&0.949 \\
\hline
\end{tabular}
}
\caption{The wave function overlaps between the Coulomb ground states at three characteristic pseudomomentum sectors and the candidate model states of the Pfaffian $|pf \rangle$, anti-Pfaffian $|apf \rangle$ and PH-symmetrized Pfaffian $|spf \rangle$ for an $N_e=12$ system on the torus with a square unit cell at the parameter set ($B=9.6$ T, $W=5$ nm).}\label{tableoverlap}
\end{table}

In Table \ref{tableoverlap}, we list the numerical results for a representative system with the square unit cell but different electron numbers. The ground states ($|GS\rangle$) of the screened Coulomb potential at three characteristic sectors are all investigated. The system ground state has exactly equal overlap with a Pfaffian state and with an anti-Pfaffian state but the value is no more than 0.80. In contrary, the wave function overlap between the Coulomb ground state and the PH-symmetrized state is generally larger than 0.94. With such high (nearly unitary) overlap, it is convincible to take the $|spf \rangle$ as an effective model state for the strongly LL-mixed incompressible phase.

\begin{figure}[htp]
\centering
\includegraphics[width=8.6cm]{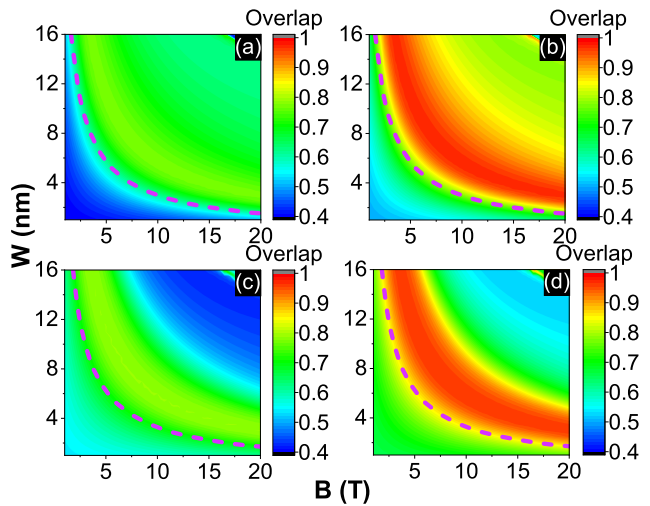}
\caption{(Color online) The wave function overlap between the Coulomb ground state and a candidate model state distributes in the ($B$,$W$) parameter space for an $N_e=12$ system on the torus with a square unit cell. The plots (a) and (b) are at (0,6) pseudomomentum sector while the plots (c) and (d) at (6,6) sector. The candidate state in (a) and (c) is a Pfaffian state while the candidate state in (b) and (d) is the PH-symmetrized Pfaffian state. Dashed curves here mark the compressible-incompressible phase boundaries as in Fig. 2.}\label{fig4}
\end{figure}

We get a more general view as we extend such numerical comparison in the full investigated parameter space. The result for an example $N_e=12$ system with a square unit cell is shown in Fig. \ref{fig4}. In the incompressible region, the symmetrized Pfaffian state at each characteristic sector exhibits a larger wave function overlap with the Coulomb ground state than a sole Pfaffian (or anti-Pfaffian) state. It is also notable in Figs. \ref{fig4}(b) and \ref{fig4}(d) that the most distributive zone (with overlaps larger than 0.90 \added{and up to 0.973}) of the symmetrized Pfaffian state is consistent with the emerging phase transition region of $\kappa_{\text{eff}}\in [2,3]$. This reassures our supposition of the incompressible state with large LL-mixing as a PH-symmetrized Pfaffian state.

Thus far, one important question is raised: if the topology of the Pfaffian
state and the anti-Pfaffian state is canceled in such a symmetrized Pfaffian
state in which the Pfaffian and anti-Pfaffian states have the same weight. If
so, the ground state of the 5/2 FQHE in ZnO, not like the PH-Pfaffian state
\cite{son}, is then topological trivial and the nonabelian excitation
disappears. To explore the topological properties of such an incompressible
state, which is close to the symmetrized Pfaffian state, we calculate its
entanglement spectrum.

\subsection{Entanglement spectrum}

The entanglement spectrum is composed of ``levels'' $\xi$ which are obtained by
the singular value decomposition of the matrix form of the ground state and are related to the
entanglement entropy \cite{entangle}. To compute the entanglement spectrum of
the ground state, we need to divide the many-body state into two parts, block
$A$ and block $B$. For completeness, we work in both the spherical and toroidal
geometries.

\begin{figure}[htp]
\centering
\includegraphics[width=8.8cm]{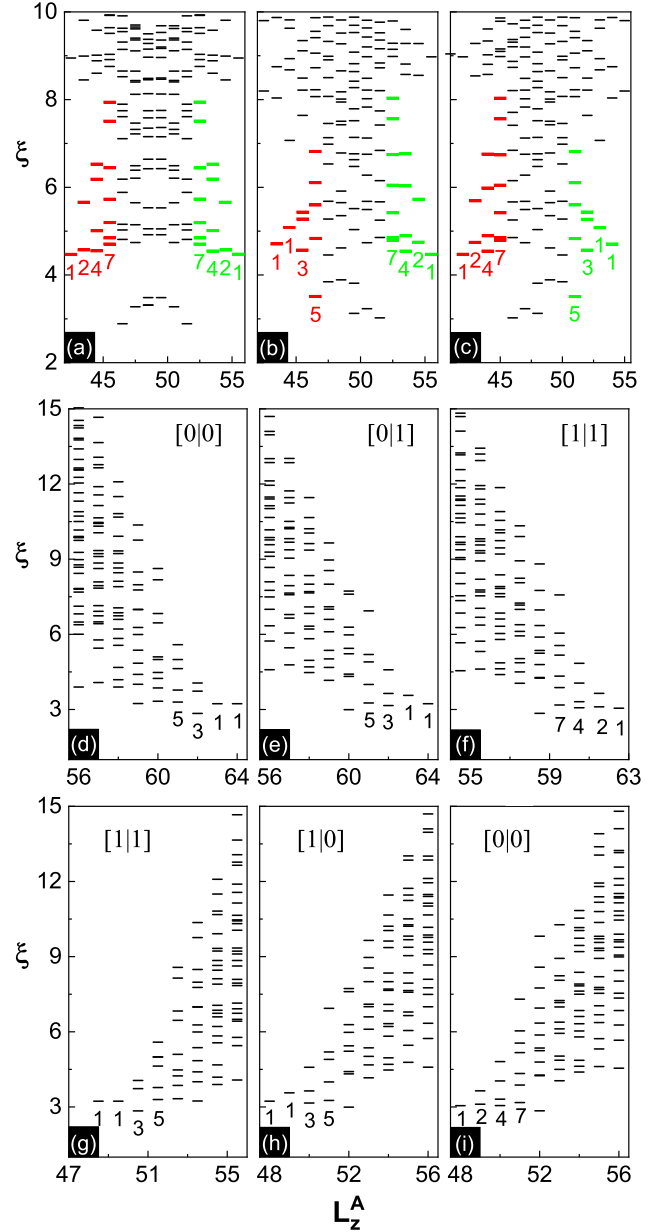}
\caption{(Color online) (a)-(c) Low-level entanglement spectra of the Coulomb ground state for an $N_e=14$ system on the sphere with the shift $S=1$ at different bipartition sets of (a)
$(N_A,N_{orb}^A)=(N_B,N_{orb}^B) =(7, 14)$, (b) $(N_A,N_{orb}^A)=(6, 13)$, and (c) $(N_A,N_{orb}^A)=(7, 13)$. In comparison, the entanglement spectra of a Pfaffian state with $N_e=16$ and $S=3$ are shown in (d)-(f) for different bipartition cuttings based on the root configuration of the Pfaffian state \cite{entangle}.  As the PH-conjugate counterpart, the entanglement spectra of an anti-Pfaffian state with $N_e=14$ and $S=-1$ are also shown in corresponding panels (g)-(i) for references. In all cases, the gapped lower levels at spectrum edges exhibit special counting patterns as labeled in the plots.} \label{fig5}
\end{figure}

In the spherical geometry, based on the previous Hall viscosity calculation, we take the shift $S=1$ to calculate the screened Coulomb system at the selected parameter set. The obtained spectrum reveals the system is incompressible with a non-degenerate ground state gapped from other higher energy states. In calculating the entanglement spectrum of this incompressible ground state, we set the bipartition block $A$ mostly locates on the semisphere with positive $z$-component of the angular momentum. The electron numbers of two partition blocks are $N_{A(B)}$ and it requires that $N_A+N_B=N_e$. The orbit numbers of two blocks are $N_{orb}^{A(B)}$ with $N_{orb}^A+N_{orb}^B =N_s$. The $z$-components of the total angular momenta of two blocks are $L_z^{A(b)}$ with $L_z^A + L_z^B=0$. 

The low-level entanglement spectra of different divisions for an example $N_e=14$ system are drawn in Fig. \ref{fig5}(a)-\ref{fig5}(c). These spectra show a two-side structure and at the edge of each side there are several lower levels gapped from higher levels. These gapped lower levels exhibit certain patterns, such as {1,1,3,5,...} or {1,2,4,7,...}, when one counts them along the $L_z^A$ decreasing (increasing) direction at the large (small) $L_z^A$ side. These unique edge-level counting patterns match with the ``fingerprint'' feature of the a Pfaffian or an anti-Pfaffian state according to the conformal field theory (CFT). As a reference, we also draw the entanglement spectra of a Pfaffian state (with $N_e=16$ and shift $S=3$) and its PH-conjugate state, the anti-Pfaffian state (with $N_e=14$ and shift $S=-1$), at different root configuration cutting edges in Figs. \ref{fig5}(d)-\ref{fig5}(f) and Figs. \ref{fig5}(g)-\ref{fig5}(i), respectively. There, the spectra of the (anti-)Pfaffian exhibit one-side gapped edge with the fingerprint patterns counted decreasingly for Pfaffian and increasingly for anti-Pfaffian. Thus, the incompressible Coulomb state owns topological feature of both the Pfaffian and anti-Pfaffian state. However, we could not straightly construct a combination state of a Pfaffian and an anti-Pfaffian on the sphere since these two states have different spherical shifts. Consequently, we could not go further to check the agreement between the screened Coulomb state and the PH-symmtrized Pfaffian state in spherical geometry. The comparison between them can be achieved in the toroidal geometry.

\begin{figure}[htp]
  \centering
  \includegraphics[width=8.7cm]{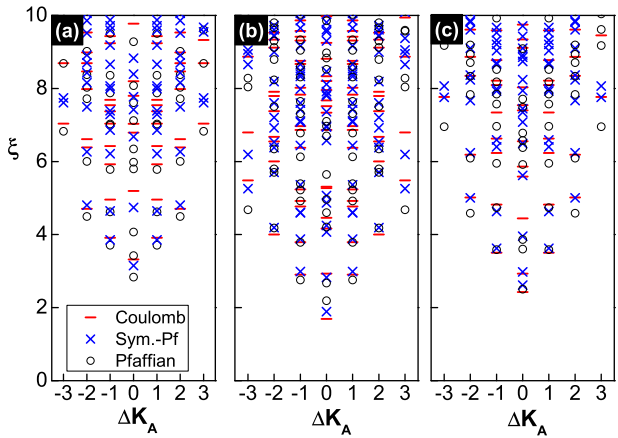}
  \caption{(Color online) The entanglement spectra of the Coulomb ground state, Pfaffian state, and PH-symmetrized Pfaffian state at the pseudomomentum sectors (a) (0,6), (b) (6,0), and (c) (6,6) for an $N_e=12$ system on the torus with a square unit cell. The bipartition blocks are evenly cut with $N_{A(B)}=6$ and $N_{orb}^{A(B)}=12$. Only the major tower around $|\Delta K_a|<4$ are shown for a better comparison.}
\label{estor}
\end{figure}

On the torus, the entanglement spectrum of a FQHE state shows a tower-like structure \cite{torusES}, which is quite different from the spherical case since herein each bipartition block contains two chiral edges. In the thin-torus limit, the entire entanglement spectrum can be understood as a direct supposition of the two nearly independent CFT edges since they are far away from each other. However, as the toroidal geometry deviates from this limit, e.g. with a square unit cell, two CFT edges get more correlated due to a short distance. In this case, the spectrum becomes more complex, in lack of a simple physical description. Nevertheless, one can probe the topological property of a FQHE state with the toroidal entanglement spectra by comparing its low-lying levels to those candidate model states.

In Fig. \ref{estor}, we exhibit the entanglement spectrum of the ground states,
Pfaffian state and PH-symmetrized Pfaffian state at $(N_e/2,0)$, $(0,N_e/2)$, and
$(N_e/2,N_e/2)$ sectors for $N_e=12$ system with each block containing $N_e/2$
electrons and $N_s/2$ orbits. Smaller systems provide with similar results. At
each sector, the spectrum levels are calculated according to the translational
momentum $K_A$, which is a sum (mod $N_s$) of the orbit indices for all
occupied electrons in the block $A$. In addition, we note that the Pfaffian and the
anti-Pfaffian states own the same edge environment due to the PH
symmetry, thus sharing the identical entanglement spectrum. We focus on the major tower structure
around the minimum level of which the momentum is shifted to be $K_A=0$. It clearly shows that the low-lying levels of the ground state entanglement spectrum closely match 
with those of the symmetrized Pfaffian state both in their level positions and patterns
while there are some mismatching with the sole Pfaffian (anti-Pfaffian) state,
indicating that the sample $a$ would rather share the same topological
universality class with the symmetrized Pfaffian state.

The above analysis reassures that the strongly-screened Coulomb state can be well described by a PH-symmetrized Pfaffian state and is confirmed to carry nontrivial dual topological nature of the Pfaffian and anit-Pfaffian states.

\section{Summary}

In this paper, we study the strongly LL-mixed 5/2 fractional quantum Hall system with even number of electrons and a finite width of the quantum well by adapting a screened Coulomb interaction as the effective Hamiltonian. The calculated low-lying spectra reveal a three-fold manifold feature in certain region of the parameter space made of the magnetic field and quantum well width. With the PCA on the wave functions of these manifold states, we detect a compressible-incompressible phase transition in the parameter space by the competition between the first two leading PCs, which is consistent with the ZnO experiments and early works on an odd-electron system. The significant presence of a subdominant third PC inside the incompressible region suggests an additional transition as the LL-mixing grows into a certain strength range, which agrees with the discovery in some early experiments and numerical works with zero-width quantum well. Further calculations on this emerging phase with the Hall viscosity and wave function overlaps show its ground states own PH symmetry and match well with the PH-symmetrized Pfaffian state rather than a sole PH-nonsymmetric Pfaffian or anti-Pfaffian state. The investigation of the entanglement spectra in both spherical and toroidal geometries demonstrates this strongly LL-mixed phase carries the nontrivial topological properties with dual features of the Pfaffian and anti-Pfaffian states.

Our results show that the $5/2$ FQHE experimentally observed in ZnO system has different topological property from the conventional GaAs system which is believed to be governed by the (anti-)Pfaffian state. Such a strongly LL-mixed FQHE state is another favorable plateau to utilize the even denominator FQHE \cite{nonab}, of which non-Abelian excitations are expected. In a thermal transport experiment, two majorana modes with half-integer Hall thermal conductance are expected. Although the two channels are in the opposite directions and may be canceled overall, in
a partial equilibrated transport \cite{nonequilibrum} the two different modes may be distinguished to some extent in ZnO system.

\section{Acknowledgment}
W.L. is supported by the NSF-China under Grant No. 11804396.  H.W. is supported by the Guangdong Provincial Key Laboratory (Grant No. 2019B121203002). We are grateful to the High Performance Computing Center of Central South University for partial support of this work.

\end{document}